\documentclass[aps,prb,twocolumn,superscriptaddress,show pacs]{revtex4}

\usepackage{amssymb}
\usepackage{amsmath}
\usepackage{graphicx}
\usepackage{natbib}
\usepackage{nicefrac}
\usepackage{multirow}
\usepackage{epstopdf}
\usepackage{float}
\usepackage{color} 

\newcommand{\ra}{\rangle}

\newcommand{\ket}[1] { |  #1 \ra }

\newcommand{\unit}[1]{\mathrm{\ #1}}

\begin{document}

\title{Hyperfine-Enhanced Gyromagnetic Ratio of a Nuclear Spin in Diamond}
\author{S. Sangtawesin}
\affiliation{Department of Physics, Princeton University, Princeton, NJ 08544, USA}
\author{J. R. Petta}
\affiliation{Department of Physics, Princeton University, Princeton, NJ 08544, USA}
\affiliation{Department of Physics, University of California, Santa Barbara, CA 93106, USA}

\begin{abstract}
Nuclear spins in the solid state environment of diamond are highly coherent, but difficult to rapidly control due to the small nuclear gyromagnetic ratio. Here we demonstrate a more than 50-fold enhancement of the effective nuclear gyromagnetic ratio by coupling the nuclear spin to an electronic spin of a nitrogen-vacancy (NV) center in diamond. The enhancement allows for faster nuclear spin rotations and is in good agreement with second-order perturbation theory. The method may be applied to other systems with similar electron-nuclear spin interactions, such as phosphorous donors in silicon, opening up the possibility of fast and direct nuclear spin control in coupled spin systems.
\end{abstract}

\maketitle
Advancements in quantum computation have been accelerated by the development of qubits with long quantum coherence times and high fidelity quantum control \cite{Barends_Nature_2014,DiCarlo:2010aa}. The nuclear spin degree of freedom has long been considered to be a good candidate for a qubit \cite{Kane:1998aa,Pla_PhysRevLett.113.246801,Muhonen:2014aa}. Several approaches towards solid state quantum computation rely on using the nuclear spin as a memory and impressive coherence times have been achieved \cite{Maurer_Science_336_2012,Dutt_Science_316_2007,Dreau_PhysRevLett.110.060502,Fuchs_NaturePhys_7_2011,Zhong_Nature2015,Saeedi_15112013}. However, many of these approaches rely on using an auxiliary qubit for manipulation of the quantum state, as direct control of the nuclear spin is typically limited to 10--50 $\mu$s Rabi periods due to the small nuclear gyromagnetic ratio \cite{Saeedi_15112013,Pla:2013aa}. In comparison, electron spin Rabi periods as short as 4~ns have been achieved in a variety of condensed matter systems \cite{Koppens:2006aa,Petersson:2012aa,vandenBerg_PRL.110.066806,Fuchs_Science_2009}.

Recent experiments on nitrogen-vacancy (NV) centers have shown that nuclear spins coupled to a NV center can exhibit Rabi oscillations that are significantly faster than expected for a bare nuclear spin that is coupled to an ac magnetic field \cite{Smeltzer_PhysRevA.80.050302, Sangtawesin_PhysRevLett.113.020506}. The faster rotation rate can be interpreted as an enhancement of the effective nuclear gyromagnetic ratio that results from hyperfine interactions with the NV electronic spin.

In this Rapid Communication, we show that the enhancement of the nuclear spin gyromagnetic ratio is tunable by varying the electronic spin transition frequency. We demonstrate over a factor of 50 improvement in the nuclear spin rotation rate compared to what is expected from the bare nuclear gyromagnetic ratio \cite{Smeltzer_PhysRevA.80.050302}. The enhancement is observed by directly measuring the nuclear spin Rabi frequency and is in good agreement with second-order perturbation theory \cite{Maze_Thesis}. The enhanced nuclear spin Rabi frequency may allow nuclear spins to be utilized in quantum control gate sequences, in addition to being a resource for a long-lived quantum memory.

\begin{figure}
\centering
\includegraphics[width=\columnwidth]{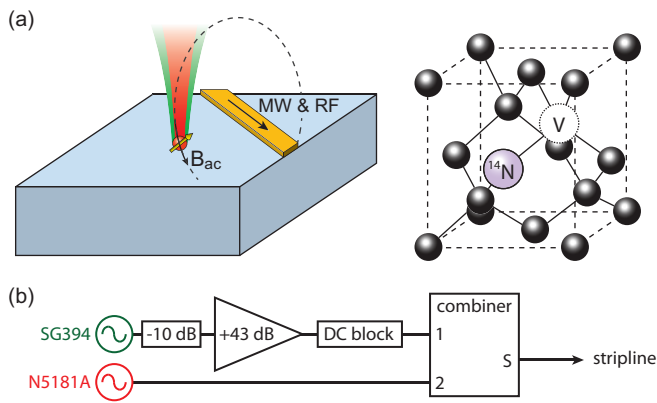}
\caption{Experimental setup: (a) A confocal microscope is used to access a single naturally occurring NV center in high-purity diamond. The NV electronic spin is optically initialized and read out with a 532 nm laser, while the intrinsic $^{14}$N of the NV is used as the nuclear spin qubit. Quantum control is achieved by applying MW and RF pulses through the short-terminated stripline fabricated on the surface of the diamond. (b) Circuit diagram: Two signal generators are used to generate MW and RF pulses. The signals are combined with an adder before being delivered to the stripline.}
\label{fig:1}
\end{figure}

The quantum system consists of a NV electron spin and its intrinsic $^{14}$N $I$ = 1 nuclear spin. A naturally occurring NV center is selected from a high-purity type-IIa diamond sample (Element Six). The experimental setup is schematically illustrated in Fig.\ \ref{fig:1}(a). A suitable NV center is located using a confocal microscope with a 532 nm excitation laser. Photoluminescence (PL) from the NV center is collected using a high numerical aperture (NA = 0.95) objective and directed towards single photon detectors using a combination of fiber and free space optics \cite{Sangtawesin_PhysRevLett.113.020506}. An ac magnetic field with amplitude $B_\mathrm{ac}$ is generated using a stripline that is fabricated on the surface of the diamond. Microwave (MW) and radio frequency (RF) pulses are applied to the stripline to drive electronic and nuclear spin rotations. The circuit diagram is shown in Fig.\ \ref{fig:1}(b). A SRS SG394 (Aglient N5181A) signal generator is used to generate MW (RF) pulses. The MW signal is amplified with a broadband amplifier (Triad RF TA1003) to allow for fast manipulation of the electronic spin. RF and MW signals are combined with a resistive splitter-combiner before they are delivered to the stripline. By driving electron Rabi oscillations and measuring the resulting Rabi frequency, we estimate $B_\mathrm{RF} \sim 2.9 \unit{G}$ at the sample with $1\unit{mW}$ of power applied to the stripline. We verify the accuracy of this conversion across the range of frequencies and powers used in this experiment by performing an independent multi-frequency excitation\cite{Childress_PRA82.033839} on the electronic spin (data not shown). By simultaneously driving the NV with RF and MW pulses, the slowly varying RF field periodically shifts the NV electronic levels, resulting in a broadened MW transition. From the width of the MW transition, we can extract the value of $B_\mathrm{RF}$ using the well-established electron $g$-factor\cite{Felton_PhysRevB.79.075203} $g_e = 2$.
\begin{figure}
	\centering
	\includegraphics[width=\columnwidth]{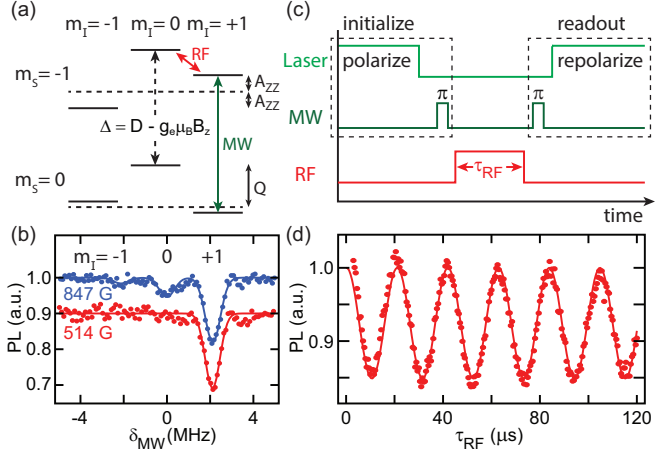}
	\caption{(a) NV center energy level diagram. A selective MW pulse is tuned to resonance with the $\ket{m_S,m_I}$ = $\ket{0,+1}$ to $\ket{-1,+1}$ transition for initialization and readout of the nuclear spin. (b) PL as a function of $\delta_\mathrm{MW} = \nu_\mathrm{MW} - \Delta$, showing nearly full nuclear spin polarization in the $m_I = +1$ level at the ESLAC (lower trace, $B_z = 514 \unit{G}$). In comparison, the polarization is significantly reduced away from the ESLAC (upper trace, $B_z = 847 \unit{G}$). The detuning $\delta_\mathrm{MW}$ is defined relative to the center frequencies $1429.7 \unit{MHz}$ ($496.4 \unit{MHz}$) for $B_z$ = 514 G ($B_z$ = 847 G). Data sets are vertically offset for clarity. (c) Pulse sequence used for nuclear spin Rabi oscillations (see text for details). (d) PL as a function of RF pulse duration $\tau_\mathrm{RF}$ showing nuclear spin Rabi oscillations.}
	\label{fig:2}
\end{figure}

The NV center Hamiltonian $H$ = $H_0$ + $V$ results in the level diagram shown in Fig.\ 2(a) and consists of secular terms $H_0$, and non-secular terms $V$:

\begin{align}
	H_0/h &= D S_z^2 + g_e \mu_B B_z S_z - g_N \mu_N \vec{B} \cdot \vec{I} \notag\\
&\hspace{3cm} - \sum_{j=x,y,z} S_z A_{zj} I_j + Q I_z^2,\label{eq:nv_hamiltonian_H0} \\
	V/h &= \frac{1}{2} \bigg(g_e \mu_B(B_- S_+ + B_+ S_-) \notag\\
	&\hspace{3cm} - \sum_{j=x,y,z} (S_+ A_{-j} I_j + S_- A_{+j} I_j)\bigg). \label{eq:nv_hamiltonian_V}
\end{align}
Here $g_e$ ($g_N$) is the electronic (nuclear) g-factor, $\mu_B$ ($\mu_N$) is the Bohr (nuclear) magneton, $\vec{B}$ is the external magnetic field, and $\vec{S}$ ($\vec{I}$) are the electron (nuclear) spin operators. Hyperfine couplings are given by the matrix components $A_{ij}$, where $i,j = \{x,y,z\}$. Raising and lowering operators are defined as $B_\pm = B_x \pm i B_y$, $S_\pm = S_x \pm i S_y$, $A_{\pm j} = A_{xj} \pm i A_{yj}$. We neglect interactions with nearby $^{13}$C impurities as their interaction strengths are much smaller than the coupling to the intrinsic $^{14}$N and no such coupling is visible in the spectroscopy data [Fig.\ \ref{fig:2}(b)]. Electronic spin states $m_S = 0$ and $m_S = -1$ are separated by $\Delta = D - g_e\mu_B B_z$, where $D = 2.87\unit{GHz}$ is the ground state zero-field splitting and $g_e \mu_B B_z$ is the Zeeman shift from the external magnetic field $B_z$ applied along the NV symmetry axis ($g_e\mu_B = 2.802 \unit{MHz/G}$). The $m_S=+1$ spin state is far detuned and therefore not shown in Fig.\ \ref{fig:2}(a). For each electronic spin state, the energy levels are further split into three sublevels by the $^{14}$N quadrupole coupling $Q = -4.96 \unit{MHz}$ and nuclear Zeeman shift ($g_N \mu_N = 0.308 \unit{kHz/G}$). Electron-nuclear axial hyperfine coupling $A_{zz} = -2.16 \unit{MHz}$ further separates the $m_I = -1, +1$ states in the $m_S = -1$ subspace. We note that while the perpendicular hyperfine terms $A_\perp = A_{xx} = A_{yy} = -2.70 \unit{MHz}$ are non-zero, the electron-nuclear flip-flop process is suppressed due to large quadrupole coupling $Q$. The off-diagonal terms of the matrix $A$ are zero due to the symmetry of the NV center \cite{Felton_PhysRevB.79.075203,Fuchs_NaturePhys_7_2011,Smeltzer_PhysRevA.80.050302}.

We select the two well-isolated sublevels $\ket{m_S,m_I} = \ket{-1,+1}$, $\ket{-1,0}$ for the nuclear spin qubit to allow for selective excitation using the RF field (frequency $\nu_\mathrm{RF}\sim 3\unit{MHz}$). This choice of states for our qubit also provides a convenient means of nuclear spin readout. We can directly map the nuclear spin state to the electronic spin state by applying a selective microwave $\pi$-pulse (with frequency $\nu_\mathrm{MW}$ tuned to resonance with $m_I=+1$ transition) at the end of the nuclear control sequence. The electronic spin state is then read out by optical excitation \cite{Dutt_Science_316_2007,Smeltzer_PhysRevA.80.050302}.

For the experiment, $B_z$ is aligned to within $\pm 1$ degree of the NV-axis \cite{Epstein_NatPhys_1_2005}. We choose to work close to the excited state level anti-crossing (ESLAC), $B_z \sim 500 \unit{G}$, to allow for efficient polarization of the nuclear spin to the $m_I=+1$ state by optical pumping with a 532 nm laser \cite{Fuchs_PhysRevLett.101.117601, Jacques_PhysRevLett.102.057403}. To probe the electronic spin transition frequencies, we apply pulsed microwave excitation with varying frequency $\nu_\mathrm{MW}$. When $\nu_\mathrm{MW}$ is on resonance with an electronic transition we observe a dip in the PL intensity. Figure \ref{fig:2}(b) shows the PL as a function of $\nu_\mathrm{MW}$, showing the transition frequencies associated with different nuclear spin projections \cite{Jacques_PhysRevB.84.195204}. Two data sets are shown for comparison. At the ESLAC ($B_z = 514 \unit{G}$), the nuclear spin is fully polarized to $m_S = +1$ (polarization $P>95\%$). Away from the ESLAC the polarization is less efficient,  as shown in the data with $B_z = 847 \unit{G}$ ($P \sim 72\%$), where transitions corresponding to $m_I = -1, 0$ can also be observed. While imperfect polarization reduces the contrast of the nuclear spin Rabi oscillations, it does not affect the measurement of the Rabi frequency and the results that follow.

The full experimental pulse sequence is illustrated in Fig.\ \ref{fig:2}(c). We start by optically pumping the electron and nuclear spins with a 4-$\mu s$ long 532 nm laser pulse. After the system is polarized into the $\ket{0,+1}$ state, a selective MW $\pi$-pulse is applied to transfer the population to the $\ket{-1,+1}$ state, completing the initialization process. We then drive nuclear spin Rabi oscillations by applying a RF pulse with varying duration $\tau_{\mathrm{RF}}$ resonant with the $\ket{-1,+1}$ to $\ket{-1,0}$ transition. Finally, optical readout is performed by applying another selective MW $\pi$-pulse that converts the population from $\ket{-1,+1}$ to the bright state $\ket{0,+1}$. This yields a PL signal that is proportional to the $\ket{-1,+1}$ population at the end of the pulse sequence.

Nuclear Rabi oscillations are readily observed by implementing this pulse sequence and plotting the PL intensity as a function of $\tau_\mathrm{RF}$. Typical nuclear spin Rabi oscillations with $B_\mathrm{RF} = 12.5 \unit{G}$ are shown in Fig.\ \ref{fig:2}(d). The nuclear Rabi frequency $\Omega_N = 48 \unit{kHz}$ we extract from the data exceeds the expected value $\Omega_N$ = $g_N \mu_N B_\mathrm{RF} = 3.85 \unit{kHz}$ by more than a factor of 10, indicating an enhancement of the effective nuclear gyromagnetic ratio.

We further examine the enhancement by plotting the nuclear Rabi frequency $\Omega_N$ as a function of $B_\mathrm{RF}$ for a series of dc magnetic fields $B_z$ (see Fig.\ 3). For each value of $B_z$, $\Omega_N$ scales linearly with $B_\mathrm{RF}$, with the observed slope implying a constant effective gyromagnetic ratio $\gamma_{N,\rm eff}$ whose value is greater than the bare nuclear gyromagnetic ratio $\gamma_{N}$ = $g_N\mu_N = 0.308 \unit{kHz/G}$. The data also confirms the dc magnetic field is well aligned with the NV axis, as there is no visible offset at $B_\mathrm{RF} = 0$ that would result from an off-axis magnetic field $B_{x,y}$. The enhancement factor $\gamma_{N,\rm eff} / \gamma_{N} = 18$ at $B_z$ = 514 $\unit{G}$ and increases to 57 at $B_z$ = 873 $\unit{G}$.

\begin{figure}
	\centering
	\includegraphics[width=\columnwidth]{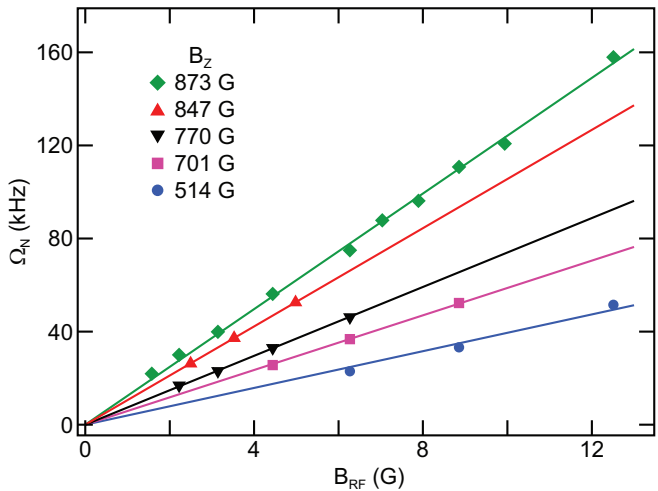}
	\caption{Nuclear spin Rabi frequency $\Omega_N$ plotted as a function of $B_\mathrm{RF}$ for several different values of $B_z$. The slope of each line corresponds to the effective nuclear gyromagnetic ratio $\gamma_{N,\rm{eff}}$. For the largest value of $B_z$ = 873 $\unit{G}$, $\gamma_{N,\rm{eff}}/\gamma_N$ = 57.}
	\label{fig:3}
\end{figure}

The enhancement of the effective nuclear gyromagnetic ratio can be understood by examining the secular and non-secular terms in the Hamiltonian, which are given by Eq.\ \ref{eq:nv_hamiltonian_H0} and Eq.~\ref{eq:nv_hamiltonian_V}, respectively \cite{Maze_PhysRevB.78.094303,Maze_Thesis}. Treating $V$ as a perturbation, and calculating to second order in perturbation theory, we have:

\begin{align}
E_{m_S}^{(2)}/h &= \frac{(3 m_S^2 - 2) D - m_S g_e \mu_B B_z}{2(D^2 - g_e^2 \mu_B^2 B_z^2)} ( \hat{M} + \hat{N} ) ,\label{eq:e2ms2}
\end{align}
where
\begin{align}
\hat{M} &= - g_e \mu_B (B_- \vec{A}_+ \cdot \vec{I} + B_+ \vec{A}_- \cdot \vec{I} ),
\end{align}
and
\begin{align}
\hat{N} &= g_e^2 \mu_B^2 B_\perp^2 + 2\vec{A}_+\vec{A}_- + (\vec{A}_+ \times \vec{A}_-)\cdot \vec{I} .
\end{align}

\noindent Here $\vec{A}_j$ = $(A_{jx},A_{jy},A_{jz})$ and $\vec{A}_\pm = \vec{A}_x \pm i \vec{A}_y$. We can neglect all the terms in $\hat{N}$ as they commute with $I_x$ and $I_y$ and therefore do not contribute to spin flips. The only remaining term, given in Eq.~\ref{eq:perturbation}, contributes directly to the nuclear spin Rabi frequency and the effective nuclear gyromagnetic ratio. For $m_S = -1$, we have,

\begin{align}
E_{m_S=-1}^{(2)}/h &= -\frac{g_e\mu_B A_\perp}{D - g_e \mu_B B_z} (B_x I_x + B_y I_y). \label{eq:perturbation}
\end{align}
\noindent The nuclear Rabi frequency is then given by:

\begin{align}
\Omega_N &= \left(g_N \mu_N - \frac{g_e \mu_B A_{\perp}}{D - g_e \mu_B B_z}\right)B_\mathrm{RF},
\end{align}

\noindent where $B_\mathrm{RF}$ is perpendicular to the NV axis.

It is now apparent that second order perturbation theory accounts for the enhanced effective nuclear gyromagnetic ratio observed in the experiment:

\begin{align}
\gamma_{N,\rm eff} = g_N\mu_N - \frac{g_e \mu_B A_{\perp}}{D - g_e \mu_B B_z}. \label{eq:g_eff}
\end{align}

\begin{figure}
	\centering
	\includegraphics[width=\columnwidth]{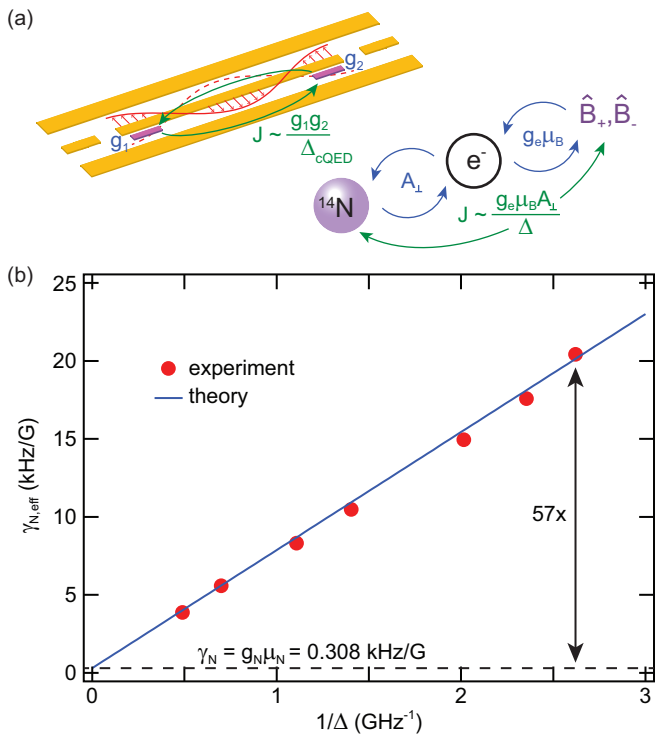}\\
	\caption{(a) Comparison of the mechanisms that lead to the cavity-mediated couplings of two qubits in the cQED architecture and the nuclear spin gyromagnetic ratio enhancement in NV centers (see text for details). (b) Effective nuclear spin gyromagnetic ratio, $\gamma_{N,\rm eff}$, plotted as a function of $1/\Delta$, and extracted from the nuclear spin Rabi data shown in Fig.~\ref{fig:3}. Enhancements $\gamma_{N,\rm eff}/\gamma_N$  $>$ 50 are obtained. The data are in good agreement with the theoretical value given by Eq.~\ref{eq:g_eff} (solid line).}
	\label{fig:4}
\end{figure}

The enhancement of the nuclear gyromagnetic ratio is analogous to two-qubit interactions in the circuit quantum electrodynamics (cQED) architecture [Fig.~\ref{fig:4}(a)] \cite{Blais_PhysRevA.75.032329,Wallraff:2004aa}. In cQED, two qubits can be coupled via the exchange of virtual photons in a microwave cavity. The effective coupling rate $J \sim g_1 g_2 / \Delta_{\rm cQED}$, is a function of the individual coupling strength of each qubit to the cavity (described by $g_1$ and $g_2$) and the qubit-cavity detuning $\Delta_{\rm cQED}$ \cite{Majer_Nature_2007}. Here our NV electronic spin acts in place of the cavity, mediating the coupling between the nuclear spin (qubit 1, coupling strength $g_1\sim A_\perp$) and the RF magnetic field (``qubit'' 2, coupling strength $g_2\sim g_e\mu_B$). The detuning is set by the energy difference $\Delta$ between the $m_S = 0$ and $m_S = -1$ states.

In Fig.\ 4(b) we plot $\gamma_{N,\rm eff}$ as a function of $1/\Delta$. The data are extracted from Fig.\ 3 and the enhancement of the effective nuclear gyromagnetic ratio is in good agreement with perturbation theory (solid line). We observe enhancements $\gamma_{N,\rm eff}/\gamma_N$ exceeding 50, making it possible to perform a full nuclear spin Rabi oscillation on a few microsecond timescale with only $\sim 10 \unit{mW}$ of power delivered to the sample. This enhancement is approximately 5 times greater than earlier results obtained on NVs at lower magnetic fields \cite{Smeltzer_PhysRevA.80.050302}. We expect the observed trends to hold for smaller energy splittings than are used in this experiment, with the trade-off being the reduction in the nuclear spin lifetime $T_{1,N}$ near the ground state level anti-crossing \cite{Dreau_PhysRevLett.110.060502}, where the energy splitting approaches zero. These smaller energy splittings are not accessible with our current experimental setup due to the difficulty of polarizing the nuclear spin far from the ESLAC.

In conclusion, we have shown that the effective nuclear spin gyromagnetic ratio can be greatly enhanced due to hyperfine coupling of the nuclear spin to the NV electronic spin. The enhancement is well described by second order perturbation theory and is analogous to cavity-mediated qubit couplings in cQED. Our approach is also applicable to other coupled electron-nuclear spin systems, such as phosphorous donors in silicon or rare-earth metal dopants in yttrium orthosilicate\cite{Pla:2013aa,Zhong_Nature2015}, where it would allow for rapid quantum control of nuclear spins without requiring high RF power.

\begin{acknowledgements}
Research was supported by the Packard Foundation, the National Science Foundation (DMR-1409556 and DMR-1420541), and the Army Research Office through PECASE award W911NF-08-1-0189.
\end{acknowledgements}

\bibliographystyle{apsrev}

\end{document}